\documentstyle[array,epsfig,twocolumn,pra,aps]{revtex}
\newcommand{\ket}[1]{|{#1}\rangle}
\newcommand{\bra}[1]{\langle{#1}|}

\begin{document}
\draft

\wideabs{

  \title{Probabilistic nonlocal gate operation via imperfect
    entanglement}
  
  \author{Jingak Jang,$^{1,2}$ Jinhyoung Lee,$^{1,2}$ M. S. Kim,$^{1}$ 
    and Y.-J. Park$^2$ }
  
  \address{$^1$ School of Mathematics and Physics, The Queen's
    University, \\Belfast, BT7 1NN, United Kingdom\\ $^2$ Department
    of Physics, Sogang University, CPO Box 1142, Seoul, Korea }
  
  \maketitle

  \date{\today}
  
\begin{abstract}  
  Nonlocal gate operation is based on sharing an ancillary pair of
  qubits in perfect entanglement. When the ancillary pair are
  partially entangled, the efficiency of the gate operation drops.
  Using general transformations, we devise probabilistic
  nonlocal gates, which perform the nonlocal operation conclusively
  when the ancillary pair are only partially entangled.  We show that
  a controlled purification protocol can be implemented by the
  probabilistic nonlocal operation.
\end{abstract}


}

\section{Introduction}
A research in quantum computation is to understand how quantum
mechanics can improve acquisition, transmission, and processing of
information. The design of any quantum computing device includes
prescriptions on how to prepare quantum memories, how to realize
quantum gate operation, and how to readout.  In quantum computation,
any quantum logic operation can be performed in a combination of
controlled-NOT (C-NOT) gates and single-bit unitary gates
\cite{DiVincenzo95}.

As a possible route toward scalable quantum computation, nonlocal
quantum gates have been suggested by Eisert {\it et al.}
\cite{Eisert00} and by Collins {\it et al.} \cite{Collins00}.  By the
nonlocal quantum operation, the phase of a qubit is changed depending
on the state of a remote qubit.  This nonlocal operation is based on
sharing an ancillary pair of qubits in perfect entanglement.  In
realizable experiments for quantum computation, it is not easy to
produce maximally entangled qubits (e-bit) so that it is worth
studying nonlocal gate operation when the ancillary e-bit is only
partially entangled.

In this paper, we devise nonlocal C-NOT gates, which perform the
operation {\it conclusively with a finite probability} when the
ancillary e-bit is {\it pure} but partially entangled.  We show that
the nonlocal gate by Eisert {\it et al.} \cite{Eisert00} can be
decomposed into two smaller units. After sharing a maximally entangled
ancillary e-bit, the first unit of operation prepares a {\it control
  e-bit} which carries the information on the control qubit. The
second unit then performs NOT operation controlled by the control
e-bit.  When the ancillary e-bit is only partially entangled, the
first unit operation becomes imperfect as the required preparation is
impossible.  We thus add an auxiliary unit of operation
between the two operations, correcting the error which occurs in the first unit.
The auxiliary unit, called the {\it corrector unit}, works conclusively 
using a general transformation. In our protocol, the general
transformation is implemented, after adding an ancillary qubit, 
either by a two-body unitary interaction and 
orthogonal measurement or by a positive operator valued measurement 
(POVM) \cite{Jauch67,Peres93}.  A POVM has been recently accomplished
in a quantum optical experiment \cite{Clarke00}.  A probabilistic
computation \cite{Grover00} may be performed using this probabilistic
nonlocal gate.

It is well-known that a C-NOT gate can maximally entangle two product
states.  When performing the probabilistic nonlocal C-NOT operation
for a partially entangled e-bit, a maximally entangled e-bit is
conclusively produced. It implies that the present protocol
concentrates entanglement from an ensemble of partially entangled
particles to a subensemble of maximally entangled ones.

\section{Nonlocal C-NOT gate}

\subsection{Partially entangled ancillary pair}
\label{sec:peap}

\begin{figure}
  \begin{center}
    \leavevmode \includegraphics[width=0.5\textwidth]{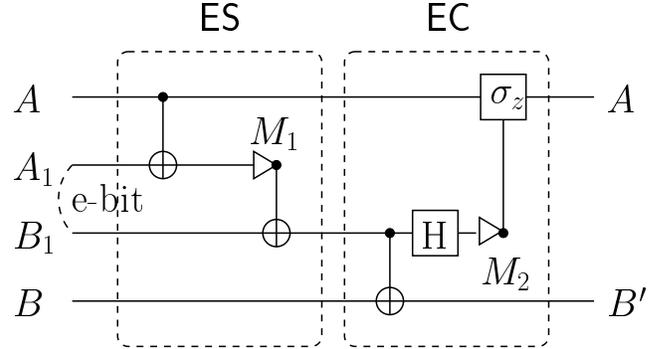}
    \caption{A nonlocal C-NOT gate for the control qubit $A$ and the
      target qubit $B$, assisted by a maximally entangled ancillary
      pair $A_1$ and $B_1$. It can be decomposed into two small units:
      ES and EC units.  The ES unit prepares the control e-bit of $A$
      and $B_1$ and the EC unit performs C-NOT-like operation between
      the control e-bit and the target bit $B$. $M_1$, $M_2$ :
      orthogonal measurements, H : Hadamard operator.}
  \label{fig:perf}
  \end{center}
\end{figure}

Before considering nonlocal C-NOT gate with partially entangled e-bit,
we discuss the main idea inherent in the protocol suggested by Eisert
{et al.} \cite{Eisert00}. The protocol is presented in Fig.
\ref{fig:perf}.  The control qubit is $A$ and the target qubit is $B$.
The ancillary e-bit, $A_1$ and $B_1$, shown in Fig.~\ref{fig:perf},
plays a crucial role in the nonlocal gate operation. Any sub-indexed
$A$ ($B$) is local to the qubit $A$ ($B$) throughout the paper.
Suppose that the control qubit $A$ is in state $\ket{A}_A = a\ket{0}_A
+ b\ket{1}_A$ and the target qubit $B$ is in $\ket{B}_B =
c\ket{0}_B+d\ket{1}_B$.  Nonlocal C-NOT operation results in
\begin{equation}
  \label{eq:nlcn}
  \ket{A}_A\ket{B}_B \rightarrow \left(ac\ket{00} + ad\ket{01} +
    bc\ket{11} + bd\ket{10}\right)_{AB}.
\end{equation}

The nonlocal C-NOT gate is composed of two smaller units.  The first
unit, shown in the left-hand-side box of Fig.~\ref{fig:perf},
entangles the control qubit $A$ to one of the ancillary e-bit $B_1$.
The ancillary e-bit prepared in
$\ket{E}_{A_1B_1}=\frac{1}{\sqrt{2}}(\ket{00}+\ket{11})_{A_1B_1}$ and
the control qubit $A$ are not entangled at the initial instance. The
local C-NOT is applied on $A$ and $A_1$ to give
\begin{eqnarray}
  & &\left(a\ket{0} + b\ket{1}\right)_A \ket{E}_{A_1B_1} \nonumber \\ 
  & &\rightarrow \frac{1}{\sqrt{2}}\left[a\left(\ket{000} +
      \ket{011}\right) + b \left(\ket{110} +
      \ket{101}\right)\right]_{AA_1B_1}.
\end{eqnarray}
The state of $A_1$ is measured and the result is transmitted to
transform $B_1$.  If the measurement outcome were $|1\rangle$, the
qubit $B_1$ is flipped.  No operation is applied otherwise.  After the
operation of the first unit the entanglement of $\ket{E}_{A_1B_1}$ is
swapped to $\ket{\Theta}_{AB_1}$:
\begin{equation}
  \label{eq:nlg}
  \left(a\ket{0} + b\ket{1}\right)_A\ket{E}_{A_1B_1} \rightarrow
  \ket{\Theta}_{AB_1} = \left(a\ket{00} + b\ket{11}\right)_{AB_1}.
\end{equation}
We call thus the first unit of the nonlocal gate as an entanglement
swap (ES) unit.  The prepared e-bit $\ket{\Theta}_{AB_1}$ carries the
quantum information of the control qubit $A$ so the e-bit
$\ket{\Theta}_{AB_1}$ is called a control e-bit.

At the second unit, the NOT operation is performed on $B$ controlled
by the control e-bit.  We thus call the second unit as the
entanglement-controlled operation (EC) unit.  In the EC unit, a local
C-NOT is applied on $B_1$ and $B$ to give
\begin{eqnarray}
  & & \left(a \ket{00} + b \ket{11} \right)_{AB_1} \left(c\ket{0} +
    d\ket{1}\right)_B \nonumber \\ & & \rightarrow \left(ac \ket{000}
    + ad \ket{001} + bc \ket{111} + bd \ket{110} \right)_{AB_1B}.
\end{eqnarray}
The $B_1$ qubit is measured after the Hadamard transformation H.  When
the measurement results in $|1\rangle$, the unitary $\hat{\sigma}_z$
is applied on the $A$ qubit. Otherwise, no operation is done on it.

Now, we consider the situation that the ancillary e-bit is in the
partially entangled pure state $\ket{\tilde{E}}_{A_1B_1} =
(\alpha\ket{00}+\beta\ket{11})_{A_1B_1}$ with $\alpha\neq\beta$ where
$\alpha$ and $\beta$ are assumed real numbers satisfying $\alpha >
\beta$.  For the partially entangled e-bit, the protocol present in
Fig.~\ref{fig:perf} does not work any longer and needs some
modification. The control e-bit produced in the ES unit depends on the
measurement outcome $m$ at the measuring device $M_1$:
\begin{eqnarray}
  \label{eq:peceb}
  \ket{\tilde{\Theta}_0}_{AB_1} &=& \frac{1}{\sqrt{p_0}} \left(a
    \alpha \ket{00} + b \beta \ket{11}\right)_{AB_1} ~~~~\mbox{for}~~~
  m = 0, \nonumber \\ \ket{\tilde{\Theta}_1}_{AB_1} &=&
  \frac{1}{\sqrt{p_1}} \left(a \beta \ket{00} + b \alpha
    \ket{11}\right)_{AB_1} ~~~~\mbox{for}~~~ m = 1,
\end{eqnarray}
where $p_0 = (a\alpha)^2+(b\beta)^2$ and $p_1 =
(a\beta)^2+(b\alpha)^2$ are the probabilities for the output $m=0$ and
$m=1$, respectively.  The output state $\ket{\tilde{\Theta}_m}_{AB_1}$
has the channel dependence of $\alpha$ and $\beta$ differently from
$\ket{\Theta}_{AB_1}$ in Eq.~(\ref{eq:nlg}).

Our task is to remove the channel dependency in the control e-bit by
adding a {\it corrector unit} in the protocol to recover the control
e-bit in the form of $\ket{\Theta}_{AB_1}$ in Eq.~(\ref{eq:nlg}).
This requires a local resource to communicate between the ES and
corrector units, which can be implemented by either local classical
communication or internal one-bit classical memory. We present two
possible protocols for the corrector unit in the following.

\subsection{Conditioned unitary operator}
\label{sec:cuo}

\begin{figure}
  \begin{center}
    \leavevmode \includegraphics[width=0.5\textwidth]{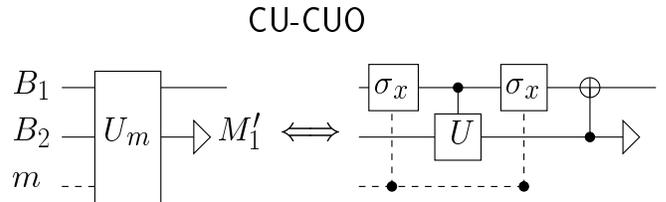}
    \caption{A corrector unit based on 
      conditioned unitary operator (CU-CUO) to prepare the
      correct control e-bit when the ancillary e-bit is only partially
      entangled.  The corrector unit is inserted between ES and EC
      units in Fig.~\ref{fig:perf}.  It works {\it conclusively} to
      make the operation free of errors. Its success is determined by
      the measurement outcome at $M_1'$. The two-qubit unitary
      operation $U_m$ can be decomposed into two
      conditioned-$\sigma_x$, a controlled-unitary $U$, and a C-NOT
      operators as shown in the right hand side.  }
  \label{fig:cut}
  \end{center}
\end{figure}

Consider a corrector unit based on conditioned unitary
operation (CU-CUO), present in Fig.~\ref{fig:cut}. The 
CU-CUO needs an ancillary qubit $B_2$ initially prepared in the
ground state $\ket{0}_{B_2}$. A two-qubit unitary
transformation is performed over qubits $B_1$ and $B_2$ \cite{Wan00},
conditioned by the measurement outcome $m$ at $M_1$. The unitary
operators $\hat{U}_1$ and $\hat{U}_2$ are given in the basis
$\{\ket{00},\ket{01},\ket{10},\ket{11}\}_{B_1B_2}$ by
\begin{equation}
\label{eq:cut1}
U_0=\left(
\begin{array}{cccc}
  \cos\theta & \sin\theta & 0 & 0 \\ 0 & 0 & 0 & 1 \\ 0 & 0 & 1 & 0 \\ 
  -\sin\theta & \cos\theta & 0 & 0
\end{array} \right)~~\mbox{for}~~m=0,
\end{equation}
and
\begin{equation}
\label{eq:cut2}
U_1=\left(
\begin{array}{cccc}
  1 & 0 & 0 & 0 \\ 0 & 0 & -\sin\theta & \cos\theta\\ 0 & 0 &
  \cos\theta & \sin\theta\\ 0 & 1 & 0 & 1 \\ 
\end{array} \right)~~\mbox{for}~~m=1,
\end{equation}
where $\cos\theta = \beta/\alpha$. For $m=0$, applying $\hat{U}_0$ on
the qubits $B_1$ and $B_2$, the composite system of $A$, $B_1$, and
$B_2$ is in the state
\begin{eqnarray}
  \ket{\tilde{\Theta}_0}_{AB_1}&& \ket{0}_{B_2}
  \stackrel{\hat{U}_0}{\longrightarrow} \nonumber \\ 
  &&\frac{1}{\sqrt{p_0}}\left( \beta \ket{\Theta}_{AB_1} \ket{0}_{B_2}
    - a \alpha \sin \theta \ket{01}_{AB_1} \ket{1}_{B_2}\right).
\end{eqnarray} 
Similarly, for $m=1$, the composite system becomes in the state
\begin{eqnarray}
  \ket{\tilde{\Theta}_1}_{AB_1}&& \ket{0}_{B_2}
  \stackrel{\hat{U}_1}{\longrightarrow} \nonumber \\ 
  &&\frac{1}{\sqrt{p_1}}\left( \beta \ket{\Theta}_{AB_1} \ket{0}_{B_2}
    - b \alpha \sin \theta \ket{10}_{AB_1} \ket{1}_{B_2}\right).
\end{eqnarray}

After the unitary transformation, the state of the qubit $B_2$ is
orthogonally measured by the measuring device $M_1'$. If
the state $\ket{0}$ is measured with the probability $\beta^2/p_m$,
the e-bit of $A$ and $B_1$ becomes in the state $\ket{\Theta}_{AB_1}$
which we want to prepare for the nonlocal C-NOT operation. If the
measurement at $M_1'$ bears the outcome $|1\rangle$, we fail the
preparation and the whole process has to be restarted.  Note that the
probability to successfully prepare the control e-bit is $2\beta^2$.

It is important to assess the  conditioned unitary
operations $\hat{U}_0$ and $\hat{U}_1$ in (\ref{eq:cut1}) and
(\ref{eq:cut2}) to see what kind of basic units we need to perform
such operations.  We find that the two-qubit unitary operators,
$\hat{U}_0$ and $\hat{U}_1$, can be decomposed into a C-NOT, a
controlled-unitary operator, and two conditioned-$\hat{\sigma}_x$
operations as shown in Fig.~{\ref{fig:cut}}.  The
conditioned-$\hat{\sigma}_x$ operator performs $\hat{\sigma}_x$
operation when the measurement outcome is $m=0$ and $\openone$ when
$m=1$.  The controlled-unitary operation is illustrated in
Table.~\ref{table:CUT}.

\subsection{Positive operator valued measurement}
\label{sec:povm}

\begin{figure}
  \begin{center}
    \leavevmode \includegraphics[width=0.35\textwidth]{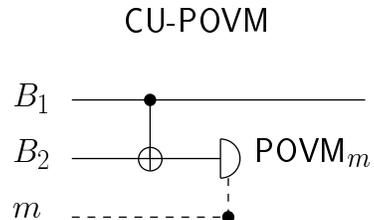}
    \caption{A corrector unit based on one-bit conditioned
      POVM (CU-POVM). The POVM$_m$ is  
      performed depending on the classical information of the
      measurement outcome $m$ from the ES unit. }
  \label{fig:povm}
  \end{center}
\end{figure}

The corrector unit can also be implemented using a 
conditioned POVM and an ancillary qubit $B_2$.  The corrector unit
based on the conditioned POVM (CU-POVM) is shown in Fig.
\ref{fig:povm}.  A set of the POVM operators is determined such that
a) the POVM operators depend on the measurement outcome $m$ on the
qubit $A_1$, b) after the measurement, we should be able to tell
either the required control e-bit, $\ket{\Theta}_{AB_1}$ is recovered
from $\ket{\tilde{\Theta}_m}_{AB_1}$, or the process has been a
failure so that we have to start again the whole operation, c) the
probability of the success is maximized.  We find the following POVM
operators satisfy the requirements:
\begin{eqnarray}
  \hat{S}_m &=& \frac{1}{\alpha^2} \ket{\psi_m}\bra{\psi_m},
\label{eq:GMOa} \\ 
\hat{F}_m &=& \openone - \hat{S}_m, \label{eq:GMOb}
\end{eqnarray}
where
\begin{equation}
\label{eq:GMO1}
\ket{\psi_0} = \beta \ket{0} + \alpha \ket{1}~~\mbox{ and}~~
\ket{\psi_1} = \alpha \ket{0} + \beta \ket{1}.
\end{equation}
Note that Eq. (\ref{eq:GMOb}) implies the completeness relation. A
straightforward algebra shows both operators being positive with
$\alpha > \beta$.

Suppose that the measurement outcome is $m=0$ at $M_1$ and the
operation of the ES unit brings about the qubits $A$ and $B_1$ in the
state $\ket{\tilde{\Theta}_0}_{AB_1}$.  An ancillary qubit $B_2$ is
initially prepared in the ground state $\ket{0}$. Applying C-NOT
operation on $B_2$ controlled by $B_1$, the composite system of $A$,
$B_1$, and $B_2$ is in
\begin{equation}
  \ket{\Psi_0} = \frac{1}{\sqrt{p_0}} (a\alpha \ket{000} + b\beta
  \ket{111})_{AB_1B_2}.
\end{equation}
The qubit $B_2$ is measured using the POVM set $\{\hat{S}_0,
\hat{F}_0\}$.  When the outcome of $\hat{S}_0$ is obtained with the
success probability of $p_s = \bra{\Psi_0} \hat{S}_0 \ket{\Psi_0} =
\beta^2/p_0$, the qubits $A$ and $B_1$ become in the state of
\begin{eqnarray}
  \label{eq:cgcnp}
  \frac{1}{p_s} \mbox{Tr}_{B_2} \hat{S}_0 \ket{\Psi_0} \bra{\Psi_0} =
  \ket{\Theta}_{AB_1}\bra{\Theta},
\end{eqnarray}
which is the required control e-bit state for the nonlocal gate. On
the other hand, if the outcome $\hat{F}_0$ is obtained, the correction
is failed and the whole operation should restart again. A similar
procedure is performed for the case of $m=1$. The POVM set is now
$\{\hat{S}_1, \hat{F}_1\}$.  When the measurement outcome is due to
$\hat{S}_m$, we get the required control e-bit.  Note that the overall
probability of the success is $2\beta^2$ which is the same as the
CU-CUO.

It is notable that instead of the POVM, an orthogonal measurement may
be employed to implement the corrector unit.  In this case, the
orthogonal measurement set is either $\{\ket{\psi_0},\ket{\phi_0}\}$
or $\{\ket{\psi_1},\ket{\phi_1}\}$ where $\ket{\psi_m}$ are defined in
Eq.~(\ref{eq:GMO1}) and $\ket{\phi_0} = \alpha \ket{0} - \beta
\ket{1}$ and $\ket{\phi_1} = \beta \ket{0} - \alpha \ket{1}$. For
either set of the orthogonal measure, the corrector unit is successful
when the state $\ket{\psi_m}$ is measured. In this case the overall
probability of success is $2\alpha^2\beta^2$, which is clearly less
than $2\beta^2$ of the CU-POVM.  Thus the CU-POVM is more optimal for
the successful operation than the corrector unit based on the
orthogonal measurement.

\subsection{Resources}
\label{sec:rs}

We have proposed two protocols for a probabilistic nonlocal C-NOT
gate.  It is useful to check the resources used in these protocols.
Here, we confine ourselves to assess the resources required by the
corrector unit.  Both protocols require a one-bit classical memory, an
ancillary qubit, a measurement, and one-bit classical communication.
The one-bit classical memory is required for communication between the
ES and corrector units because the corrector unit processes the output
state of the ES unit depending on its measurement result.  Eisert {\it
  et al}.  found that one bit of classical communication in each
direction and one shared e-bit is necessary and sufficient for the
nonlocal implementation of a quantum C-NOT gate when the e-bit is
maximally entangled \cite{Eisert00}.  When the probabilistic nonlocal
C-NOT gate operation is implemented using an partially entangled
e-bit, the operation has a probability to fail.  We have to introduce
a measurement to know the success of the operation and its measurement
result has to be communicated.  This requires an extra measurement,
and one-bit classical communication.

Comparing the two protocols in terms of required resources, it
suffices to consider the types of measurements in the 
CU-CUO and the CU-POVM.  In the CU-CUO, the one-bit
orthogonal measurement is performed.  In the CU-POVM, on the other
hand, one-bit POVM is performed so we need to expand the Hilbert space
by adding at least one extra qubit, which enables to measure
nonorthogonal states conclusively. Thus, the CU-POVM needs an
additional qubit so as to have its optimal success probability.  To
achieve the same success probability $2\beta^2$, the CU-CUO 
employs the less resources than the CU-POVM.

\section{Remarks}

One of the important properties of the C-NOT operation is to generate
or to remove the entanglement between two qubits.  Let us assume that
we initially prepare a control qubit $A$ in
$(\ket{0}\pm\ket{1})/\sqrt{2}$, a target qubit $B$ in $\ket{0}$ and a
shared e-bit which is partially entangled. After performing the
nonlocal C-NOT operation using an imperfect channel, we obtain a
maximally entangled pair. We can thus say that the shared imperfect
channel is purified to the perfect entangled channel. The optimal
probability of purification scheme via entanglement swapping is known
as $2\beta^2$ \cite{Bose99}. The probabilistic nonlocal C-NOT gate
also gives the same optimal probability. The advantage of this method
is any kind of maximally entangled pure states can be generated by
preparing accordingly the initial states of qubits $A$ and $B$.

Quantum entanglement lies in the heart of the nonlocal operation. We
have proposed the probabilistic nonlocal C-NOT gates based on
the general transformation. They have the same optimal
probabilities of success $2\beta^2$. If successful, the operation is
faithfully done and, more importantly, we know when it is faithful.
We have compared the required resources.  When the initial states are
appropriately prepared, the probabilistic nonlocal C-NOT gate in
effect refines the partially entangled state to the perfect entangled
state. This may thus serve as a purification protocol for generating
maximally entangled states.

\acknowledgements

This work was supported by the UK engineering and physical sciences
research council (EPSRC) and by Brain Korea 21 project (D-1099) of the
Korean Ministry of Education.

\begin{table}[htpb]
  \caption{Controlled-unitary operation for the control qubit $B_1$
    and the target qubit $B_2$.  $\cos{\theta}=\beta/\alpha$.}
  \begin{center}
    \begin{tabular}{|c|c||c|c|} \hline
      \multicolumn{2}{|c||}{input} & \multicolumn{2}{c|}{output} \\ 
      \hline $B_1$ & $B_2$ & $B_1$ & $B_2$ \\ \hline $\ket{0}$ &
      $\ket{0}$ & $\ket{0}$ & $\ket{0}$ \\ \hline $\ket{0}$ &
      $\ket{1}$ & $\ket{0}$ & $\ket{1}$ \\ \hline $\ket{1}$ &
      $\ket{0}$ & $\ket{1}$ &
      $\cos{\theta}\ket{0}+\sin{\theta}\ket{1}$ \\ \hline $\ket{1}$ &
      $\ket{1}$ & $\ket{1}$ &
      $-\sin{\theta}\ket{0}+\cos{\theta}\ket{1}$ \\ \hline
    \end{tabular}
  \end{center}
  \label{table:CUT}
\end{table}

\end{document}